1   **Temporal Variability of Waves at the Proton Cyclotron Frequency Upstream**

2   **from Mars: Implications for Mars Distant Hydrogen Exosphere.**

4   C. Bertucci[1,2*], N. Romanelli[1], J.Y. Chaufray[3], D. Gomez[1,2], C. Mazelle[4], M. Delva[5],

5   R. Modolo[3], F. González-Galindo[6], and D. A. Brain[7].

7   1) Instituto de Astronomía y Física del Espacio, CONICET/UBA, Buenos Aires,

8   Argentina.

9    2) Departamento de Física, FCEN, Universidad de Buenos Aires, Argentina

10  3) LATMOS, Guyancourt, France

11  4) IRAP, Toulouse, France

12  5) IWF-ÖAW, Graz, Austria

13  6) Instituto de Astrofísica de Andalucía, Granada, Spain.

14  7) University of Colorado, USA.

15  * Corresponding author (cbertucci-at-iafe.uba.ar)

17  **Abstract**

18  We report on the temporal variability of the occurrence of waves at the local proton

19  cyclotron frequency upstream from the Martian bow shock from Mars Global

20  Surveyor observations during the first aerobraking and science phasing orbit periods.

21  Observations at high southern latitudes during minimum-to-mean solar activity show

22  that the wave occurrence rate is significantly higher around perihelion/southern

23  summer solstice and lower around the same hemisphere's spring and autumn

24  equinoxes. A similar trend is observed in the hydrogen (H) exospheric density profiles

25  over the Martian South Pole obtained from a model including UV thermospheric

26   heating effects. In spite of the complexity in the ion pick-up and plasma wave

27   generation and evolution processes, these results support the idea that variations in the

28   occurrence of waves could be used to study the temporal evolution of the distant

29   Martian H corona and its coupling with the thermosphere at altitudes currently

30   inaccessible to direct measurements.

31

34

35   **1. Introduction**

36   The absence of a significant intrinsic magnetic field at Mars (Acuña et al., 1998)

37   results in the direct interaction of the magnetized Solar Wind (SW) with the planet's

38   atmosphere. Mars interaction starts far beyond the bow shock, where exospheric

39   particles get ionized. Ionization processes add a small amount of energy to the

40   newborn ions with respect to their neutral precursors. As the latter are considered to

41   be approximately at rest with respect to the planet, the planetocentric velocities of

42   newborn ions are also assumed to be negligible. The newborn ion's initial motion in

43   the SW frame consists in a gyration (ring component) around the interplanetary

44   magnetic field (IMF) and a parallel motion (beam component) at the speed of its

45   neutral precursor. The speeds of the beam and ring components are: $V_\parallel = V_{SW}$

46   $\cos(\alpha_{VB})$ and $V_\perp = V_{SW} \sin(\alpha_{VB})$ respectively, where $\alpha_{VB}$, the IMF cone angle, is the

47   initial pitch angle of the newborn ion and $V_{SW}$ is the velocity of the solar wind. The

48   newborn ion distribution function arising from the pick-up of a large number of

49   planetary ions is unstable to the growth of plasma low frequency waves (Wu and

50   Davidson, 1972). In particular, the occurrence of waves at the local cyclotron

51    frequency $\Omega_i = q_iB/m_i$ ($B$ is the magnetic field strength, and $q_i$ and $m_i$ are the charge

52    and mass of the ion respectively) of a particular ion in the SC frame can be associated

53    with the occurrence of the exospheric pick-up of ions with a specific mass-per-charge

54    ratio. This represents a potentially useful diagnostic tool to detect ionized exospheric

55    particles. An extensive discussion on the way in which magnetic field wave

56    measurements in the SC frame can be associated with ion-ion instabilities arising

57    from the pick-up of exospheric ions can be found in Romanelli et al. (2013).

58    At Mars, Phobos-2 and Mars Global Surveyor (MGS) magnetometers detected

59    upstream waves at the local proton cyclotron frequency $\Omega_p$ (Russell, et al., 1990,

60    Brain et al., 2002; Mazelle et al., 2004; Wei and Russell, 2006; Romanelli et al.,

61    2013). All cases analyzed show a left-hand polarization in the SC frame, and a quasi-

62    parallel propagation with respect to the IMF. Romanelli et al. (2013) analyzed the

63    properties of upstream waves at $\Omega_p$ from a set of 372 MGS orbits from 27 March 1998

64    to 24 September 1998, during the science phasing orbits (SPO) phase. They revealed

65    that the occurrence of waves at $\Omega_p$ strongly dropped between SPO1 (62% of the

66    upstream observation time) and SPO2 (8%) sub-phases. They also noted no

67    significant difference in the directional properties of the IMF, $\alpha_{VB}$, or the IMF's

68    convective electric field between the two sub-phases, suggesting that the reduction in

69    wave occurrence could be due to temporal changes in the density of pick-up protons.

70    Theoretical and observational studies have confirmed that the source of the pick-up

71    ions responsible for the occurrence of waves at $\Omega_p$ at Mars is its hydrogen (H)

72    exosphere via ionization processes such as photoionization and charge exchange

73    (Barabash and Lundin, 2006; Dennerl et al., 2006). As in the case of Venus, the

74    Martian (H) corona has been hypothesized to contain a hot and a relatively cooler

75    thermal population (see, e.g., Johnson et al., 2008). A recent work by Chaufray et al,

76   (2008) based on Mars Express SPICAM observations suggests the presence of hot (T

77   > 500 K) and cold (T~200 K) populations. However, the inferred hot H densities are

78   not supported by theory and the two-population model is still poorly constrained by

79   observations and model errors. Other works find inconclusive evidence for a two-

80   population model (Feldman et al., 2011) and provide single profiles.

81   So far, only models and indirect observations can predict the structure of the

82   exosphere at altitudes higher than a few scale heights above the exobase. Most models

83   use a Chamberlain (1963) approach based on isothermal equilibrium. The vertical

84   profile of the exospheric number density is an exponential whose scale height is a

85   function of the exobase temperature and density, provided by models or observations.

86   Long term (timescales larger than diurnal) variations in the modeled exospheric

87   densities come mainly from UV heating of the thermosphere. As a result, solar cycle,

88   annual and seasonal influences are expected. Simulations including a self-consistent

89   calculation of the global ion production (Modolo et al., 2005) show that H escape is

90   strongly dependent on the EUV flux through its influence on exospheric densities.

91   In this work, we study the temporal variability of the occurrence of upstream waves at

92   $\Omega_p$ at Mars based on MGS MAG observations during the first Aerobraking (AB1) and

93   SPO phases and we propose an explanation for the observed trends based on the

94   behavior of Mars distant H exosphere. Then, we discuss the implications of these

95   results in the understanding of Mars H corona and its interaction with the solar wind.

96

## 2. Observations

98   MGS MAG is a dual fluxgate magnetometer that provides fast (up to 32 Hz), wide

99   range (±4, ±65335 nT) magnetic field measurements with an uncertainty of ±1nT due

100  to spacecraft fields (Acuña et al., 2001). Typical values for the proton cyclotron

101   frequency at Mars orbit ($\sim$ 0.05 Hz) justify the use of low-resolution MAG data

102   (0.167 - 1.33 Hz) for which additional calibration is available.

103   After Mars orbit insertion/capture (MOI) on September 11, 1998, MGS was placed in

104   highly elliptical orbits with apoapses approximately located over the planet's South

105   Pole. The evolution of MGS orbital period from MOI through the entire pre-mapping

106   phase is described in Figure 6 from Albee et al. (2001). The aerobraking applied

107   between MOI and April 1998 reduced the orbital period from 48 to 12 h; this phase is

108   known as AB1. SPO came in after AB1 and ended in November 1998, when a second

109   aerobraking (AB2) phase began. SPO orbits were 12-hour long and local times

110   monotonically varying between noon and 4AM. During AB1 and SPO, the southern

111   latitudes of the apoapses remained high.

112   The upstream segments of every AB1 and SPO orbit were examined in the search for

113   waves at $\Omega_p$. Since these waves propagate almost parallel to the IMF (Romanelli et

114   al., 2013), the criterion used to identify them was based on the Fourier power spectral

115   density (PSD) of the transverse wave components with respect to the IMF ($\delta\mathbf{B}_\perp$). The

116   PSD of $\delta\mathbf{B}_\perp$ is the norm of a vector whose components are the power spectral densities

117   the components of $\delta\mathbf{B}_\perp$. Its value is estimated over sliding 10-minute intervals with an

118   overlap of 1 minute. Upstream portions are identified using the points where MGS

119   trajectory crosses Vignes et al (2000) bow shock fit, assuming an error of 10 minutes

120   on the upstream side along the spacecraft trajectory.

121   We define that waves at $\Omega_p$ are detected when the average PSD within the frequency

122   band $\Omega_p \pm 0.015$ Hz (0.015 Hz being the error in $\Omega_p$ associated with MAG's

123   uncertainty) is larger than the average PSD for frequencies $\omega > \Omega_p + 0.015$ Hz, plus

124   its standard deviation (STD), multiplied by a constant $k > 1$. If $f_N$ is the Nyquist

125  frequency (0.167 and 0.667 Hz for the two low frequency sampling rates available),

126  this is: $\langle PSD(\omega)\rangle_{\Omega p-0.015Hz}^{\Omega p+0.015Hz} > k\left\{\langle PSD(\omega)\rangle_{\Omega p+0.015Hz}^{f_N} + STD\{PSD(\omega)\}_{\Omega p+0.015Hz}^{f_N}\right\}.$

127  The value of k is obtained by inspecting several orbits with and without a clear

128  spectral line at $\Omega_p$. A value of k = 2.5 was found to be acceptable as all selected orbits

129  show signatures at $\Omega_p$, and their number is adequate for statistical purposes. Similar

130  criteria have been applied on Venus Express magnetometer measurements (Delva et

131  al., 2011, and references therein).

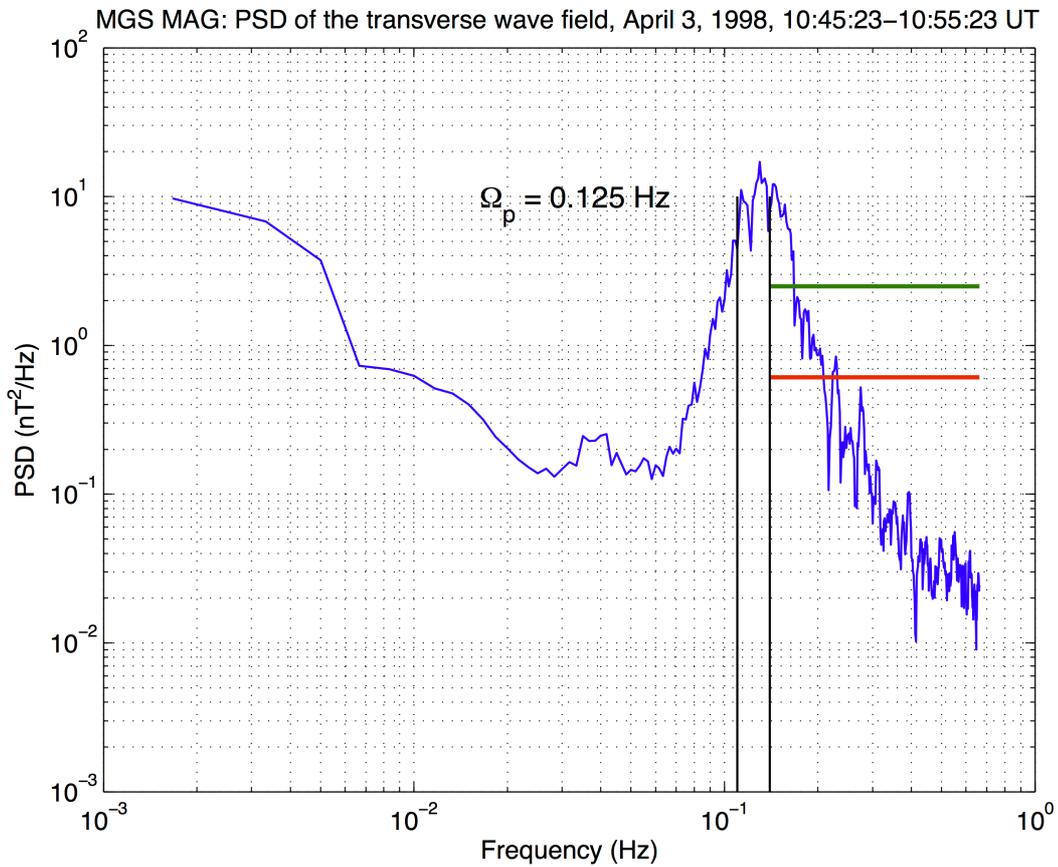

132

133  *Figure 1. PSD of  δ**B***₁ *for 10:45:23-10:55:23 on April 3, 1998 for a case of positive detection of waves*

134  *at Ω_p. Black lines indicate the frequency range corresponding to the error in the estimate of Ω_p due to*

135  *spacecraft fields (Acuña et al., 2001). The average power density for frequencies higher than the upper*

136  *limit of the interval is indicated in red. The green line shows that average plus the standard deviation.*

137  *The latter value is 3.92 times smaller than the power in the interval around Ω_p.*

138

139    Figure 1 shows an example of a positive detection of waves at $\Omega_p$. The plot shows the

140    PSD of $\delta \mathbf{B}_\perp$ as a function of frequency $\omega$ for the interval 10:45:23-10:55:23 on April

141    3, 1998. The spectrum shows a peak at $\Omega_p = 0.125$ Hz in the SC frame. The power

142    contained in the 0.105 - 0.135 Hz band (black lines) is higher than the average power

143    contained at higher frequencies (red line), plus that power's STD (green line).

144    For statistical purposes, a wave occurrence rate for each orbit was estimated as the

145    number of intervals meeting the criterion above, divided by the total number of

146    intervals in the upstream portion of the orbit. The occurrence rates are then averaged

147    every 15 orbits with a 14-orbit overlap, regardless of the orbit period. The averaged

148    occurrence rate is hereafter referred to as R.

149

## 3. Results and Discussion

151    Figure 2a shows the evolution of R between September 14, 1997 and September 17,

152    1998 (AB1, and SPO). Data gaps corresponding to the AB1-SPO1 transition

153    (February 20 - March 26, 1998), and solar conjunction hiatus (April 30 - May 26,

154    1998) are apparent. The occurrence rate exhibits an increase at the very beginning of

155    the mission -around the southern spring equinox (September 12, 1997). R reaches a

156    local maximum of 19% around October 27, 1997 and then goes back to near zero

157    around day November 16, 1997. After that, R increases again, and reaches 50% by the

158    end of the year, and a maximum at 68 % a few days after Mars perihelion (January 7,

159    1998). After January 12, the wave abundance falls again, but this time values remain

160    above 40% before the first data gap. The highest values around 89% are clustered

161    around April 1998, during southern summer. After solar conjunction, a more gradual

162    decrease is observed, with values around 20% during early southern autumn.

163    Occurrence rates are consistent with the map by Brain et al. (2002).

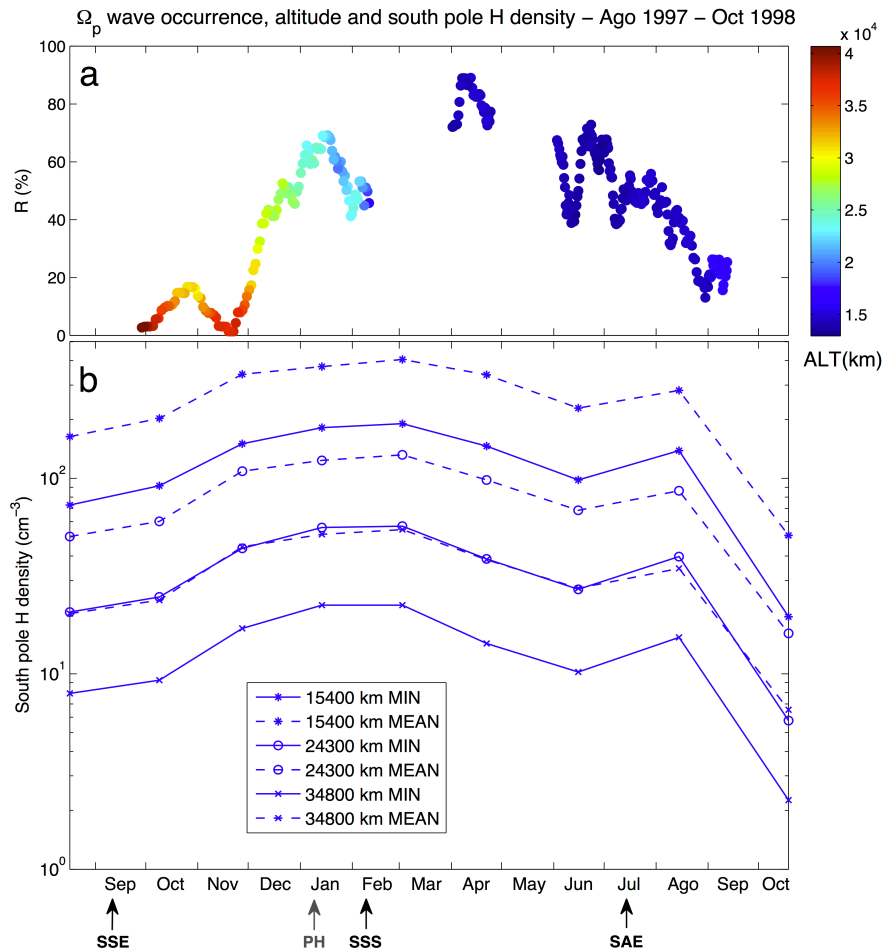

Figure 2a: Occurrence of waves at $\Omega_p$ from Sep. 1997 through Sep. 1998 averaged every 15 MGS orbits. Points are color-coded according to the upstream average altitude (see text for details). Figure 2b: Modeled exospheric H density above Mars South Pole at 15400, 24300 and 34800 km altitude for minimum and mean solar activity from Aug. 1997 to Oct. 1998. Dates for Mars perihelion (PH, January 7, 1998), southern hemisphere spring equinox (SSE, 12 September 1997), southern hemisphere summer solstice (SSS, February 6, 1998), and southern hemisphere autumn equinox (SAE, July 14, 1998) are indicated.

The changes in R can be better understood if the average altitudes of the upstream orbit portions are considered. Mean altitudes are higher during AB1 ($\sim$30000 km), where fewer wave events are observed. In particular an anti-correlation between R

176    and the average altitude is observed until January 1998. During SPO, average

177    altitudes are around 13600 km, suggesting a weaker influence of radial distance on R.

178    Figure 2a suggests a temporal dependence in the abundance of upstream waves at $\Omega_p$.

179    The hypothesis that any variability in the wave occurrence at a given location is

180    attributed to temporal effects is supported by the strong difference in wave abundance

181    for orbits with similar geometry, and the absence of a manifest influence upon the

182    convective electric field or the IMF direction (Romanelli et al., 2013). In what

183    follows, changes in the efficiency of the wave generation mechanism, the properties

184    of the solar wind, the efficiency of the ionization mechanisms and/or the local

185    exospheric H density are discussed in the context of MGS observations.

186

187    *3.1 Wave generation and evolution*

188    A central question is how representative the presence of waves at $\Omega_p$ is for the local

189    neutral density. Cowee et al. (2012) suggest that the increase in the amplitude of

190    waves at $\Omega_p$ is due to the increase in the local pick-up ion density and the cumulative

191    ion production upstream of the observation point. As exospheric densities increase,

192    increases in both the local and cumulative ion productions are expected. Accordingly,

193    higher occurrence rates will be expected at $\Omega_p$ as long as nonlinear effects do not

194    degrade the wave spectra below the detection threshold. According to Cowee et al.,

195    (2012) the observed waves would be below nonlinear saturation.

196    Another aspect to consider is that the wavelength for maximum linear growth rates at

197    least for the ion-ion right hand mode is of the order of a Martian radius (Bertucci,

198    2003), setting a limit on the spatial resolution in the use of waves as indicators of

199    exospheric densities. To avoid this limitation, comparisons between the spatial

200   distributions of waves and exospheric densities are restricted to length scales greater

201   than a planetary radius.

202

*3.2 Changes in Solar wind parameters and ionization rates*

204   Self-consistent hybrid simulations (Modolo et al., 2005) suggest that charge exchange

205   is the main source of escaping protons and that the escape rate of these protons is four

206   times higher during solar minimum with respect to solar maximum due to the

207   expansion of the neutral H corona. Although MGS measurements are not able to

208   characterize the photoionization and charge exchange rates, the solar activity during

209   the pre-mapping phase was typical of a minimum, suggesting that both ionization

210   rates would not have experienced major alterations. In agreement with this view,

211   recent works addressing the generation of the same type of waves (Cowee et al.,

212   2012) assume constant ionization rates.

213   Transient solar wind disturbances such as ICMEs have been proven to alter

214   significantly the Martian plasma environment (Crider et al., 2005). Once again, the

215   expected low solar activity, the absence of drastic short-term (a few days long)

216   changes in the wave abundance, and the fact that from September 1997 to September

217   1998 the solar F10.7 flux at Mars did not vary significantly ($107.90 \pm 18.59$ cm$^{-2}$ s$^{-1}$)

218   suggest that such events might not be the cause of the long term variability of R.

219

*3.3 Changes in the extent of the distant Martian hydrogen corona*

221   Figure 2b shows values of H exospheric density above the Martian South Pole at

222   15400, 24300 and 34800 km altitude for solar mean and minimum activity conditions

223   between August 1997 and October 1998. H densities are provided by a 3D exospheric

224   model (Chaufray et al. 2012), which incorporates non-uniform densities and 225    temperatures at the exobase from a 3D LMD-GCM model (Gonzalez-Galindo et al.

226    2009). Values initially given as a function of solar longitude are then converted to

227    time using MGS ephemeris. Three basic properties are immediately evident. First,

228    densities fall by one order of magnitude between 15400 and 34800 km for each solar

229    activity scenario. Second, H densities increase with solar activity, which reveals the

230    UV control of the thermosphere. Third, the influence of solar declination and

231    heliocentric distance is manifested in density curves displaying maxima slightly after

232    Mars perihelion and southern summer solstice (February 6, 1998), and lower values

233    during spring and autumn. H densities are not symmetric around their maxima. This is

234    probably due to heliocentric distance effects (aphelia occurred on January 29, 1997,

235    and December 16, 1998, respectively).

236    Although the wave growth rate depends on several factors, the observed variability in

237    R could be explained from MGS orbital geometry and the annual and seasonal

238    changes in the H exospheric profile predicted by models. Early in AB1, MGS

239    explored, on average, regions with altitudes around the 34800 km altitude level during

240    solar minimum. The low H densities at that altitude for solar minimum (around 10 cm$^{-}$

241    $^{3}$) could be responsible for the low wave occurrence rates observed. Furthermore, the

242    increase in R around the end of October 1997 is consistent with MGS exploring lower

243    altitudes, where higher H densities ($\sim$ 20 cm$^{-3}$) and wave amplitudes (Romanelli et al.,

244    2013) are expected. As AB1 progresses, the increase in the wave occurrence could be

245    a result of both the decrease in the average altitude of the observations and the

246    increase in the local H density as southern summer approached. The peak in

247    occurrence observed in mid January 1998 is followed by a short $\sim$20% drop which

248    cannot be explained, as predicted H densities remain high and average orbital altitude

249    continues to decrease.

250    The rather constant average altitudes during SPO allow a direct comparison of the

251    wave abundance and H density curves the lowest altitude level. Apart from the data

252    gaps, these curves display similar trends, as the southern hemisphere gradually

253    transits towards the autumn equinox (July 14, 1998). The gradual decrease in R could

254    be then due to the change in exospheric densities with solar activity. Finally, in mid

255    September, the wave abundance is around 20 % and local predicted densities are

256    closer to the values expected in mean solar activity conditions ($\sim 20$ cm$^{-3}$).

257    This comparison provides a possible explanation to the temporal variability in the

258    occurrence of waves at $\Omega_p$ based on the expansion and contraction of the Martian H

259    exosphere due to annual and seasonal components in the solar UV heating of the

260    thermosphere. If this is confirmed, the temporal variability in the occurrence of waves

261    at $\Omega_p$ could be used to study the coupling of the thermosphere with the exosphere of

262    Mars (Bougher et al., 1999), where direct observations are unavailable. In particular,

263    wave observations coincide with the occurrence of dust storms of varied intensity

264    (Keating, et al., 1998; Clancy et al., 2000) with known effects on the Martian

265    thermosphere.

266    The influence of the exosphere in the Martian solar wind interaction has also been

267    suggested to be responsible for other plasma structures such as the magnetic pileup

268    boundary (MPB, see, e.g., Crider et al., 2000). A study based on MGS MAG/ER data

269    during the mapping phase revealed that the altitude of the northern MPB (i.e., far

270    from the influence of crustal fields) is sensitive to Martian seasons (Brain et al.,

271    2005). The question whether this behavior is related to the hypotheses presented here

272    or not is beyond the scope of this work. Nevertheless, all these results suggest that in

273    addition to the solar cycle, the annual and seasonal variability in the UV irradiance

274    could be a potentially important component of the temporal evolution of the Martian

275    plasma environment.

276

## 4. Conclusions

278    The pick-up of exospheric ions, and the subsequent growth and evolution of plasma

279    waves via microscopic plasma interactions are complex physical processes. By virtue

280    of a particular property of waves propagating along the IMF and originating from

281    exospheric ion pick-up, the analysis of the occurrence of upstream transverse

282    fluctuations at $\Omega_p$ at high southern latitudes shows that the abundance of such waves

283    varies with time. This temporal variability is found to display similar trends as those

284    of the densities of exospheric H obtained from models, which take the effect of

285    thermospheric heating by solar UV radiation into account. The underlying assumption

286    is that wave generation mainly depends on the H local density and that ionization

287    rates remain constant. This approach follows previous works, which suggest that the

288    increase in the ion production is due to changes in the exospheric densities.

289    Future studies based on improved exospheric models and more comprehensive

290    measurements (such as those to be carried out by the MAVEN mission) will help to

291    confirm or discard the proposed interpretation. In particular, the effects of the vertical

292    transfer of energy and momentum at different timescales, including the role of global

293    atmospheric events such as dust storms should be investigated.

294

360

361   **Acknowledgements**

362   This work was done in conjunction with the International Space Science Institute

363   (ISSI) International Team on Comparative Induced Magnetospheres. FGG is funded

364   by a CSIC JAE-Doc contract co-financed by the European Social Fund.

365